\documentclass{svmult}

\usepackage{makeidx}         
\usepackage{graphicx}        
\usepackage{multicol}        
\usepackage[bottom]{footmisc}

\usepackage{amsmath}
\usepackage{amssymb}
\usepackage{amsfonts}
\usepackage{cite}

\usepackage{bbm}
\usepackage{ae}
\usepackage{subfigure}

\makeindex             

\newpage{\pagestyle{empty}\cleardoublepage} 

\newcommand{\bra}[1]{\langle #1|}
\newcommand{\ket}[1]{|#1\rangle}

\newcommand{\Tr}{\rm Tr\,}
\renewcommand{\Im}{\rm Im\,}
\renewcommand{\Re}{\rm Re\,}

\begin{document}

\title*{Localization of electronic states in amorphous materials: recursive Green's function method and the metal-insulator transition at $E\neq 0$}
\titlerunning{Localization of electronic states in amorphous materials\quad\hfil } 
\author{Alexander Croy\inst{1}\and
Rudolf A. R\"{o}mer\inst{2}\and Michael Schreiber\inst{1}}
\institute{Institut f\"{u}r Physik, Technische Universit\"{a}t,
09107 Chemnitz, Germany \texttt{alexander.croy@s2000.tu-chemnitz.de, schreiber@physik.tu-chemnitz.de}
\and Centre for Scientific Computing and Department of Physics,
University of Warwick, Coventry, CV4 7AL, United Kingdom
\texttt{r.roemer@warwick.ac.uk}}
%
%
\maketitle

\section{Introduction}
\label{sec:intro}

Traditionally, condensed matter physics has focused on the investigation of
perfect crystals. However, real materials usually contain impurities,
dislocations or other defects, which distort the crystal. If the deviations from
the perfect crystalline structure are large enough, one speaks of {\it disordered
systems}. The Anderson\index{Anderson} model \cite{CRAnd58} is widely used to
investigate the phenomenon of localisation\index{localisation} of electronic
states in disordered materials and electronic transport properties in
mesoscopic\index{mesoscopic} devices in general. Especially the occurrence of a
quantum phase transition driven by disorder from an insulating phase, where all
states are localised, to a metallic phase with extended states, has lead to
extensive analytical and numerical investigations of the critical\index{critical}
properties of this metal-insulator\index{insulator} transition\index{transition}
(MIT)\index{MIT} \cite{CRKraM93,CRRomS03,CRPlyRS03}. The investigation of the
behaviour close to the MIT is supported by the one-parameter scaling
hypothesis\index{one-parameter scaling hypothesis} \cite{CREndB94,CRVilRS99a}.
This scaling theory\index{scaling theory} originally formulated for the
conductance plays a crucial role in understanding the MIT \cite{CRAbrALR79}. It
is based on an ansatz interpolating between metallic and insulating regimes
\cite{CRLeeR85}. So far, scaling has been demonstrated to an astonishing degree
of accuracy by numerical studies of the Anderson model
\cite{CRSleO99a,CRMilRS00,CRMilRSU00,CRNdaRS02a,CRNdaRS04}.  However, most
studies focused on scaling of the localisation length and the conductivity at the
disorder-driven MIT\index{disorder-driven MIT} in the vicinity of the band centre
\cite{CRSleO99a,CRSleMO01,CRBraHMM01}. Assuming a power-law form for the d.c.\
conductivity,\index{conductivity} as it is expected from the one-parameter
scaling theory, Villagonzalo et al.\ \cite{CRVilRS99a} have used the
Chester-Thellung-Kubo-Greenwood formalism to calculate the temperature dependence
of the thermoelectric properties numerically and showed that all thermoelectric
quantities follow single-parameter scaling laws \cite{CRVilRS99b,CRVilRSM00a}.

In this chapter we will investigate whether the scaling
assumptions made in previous studies for the transition at
energies outside the band centre can be reconfirmed in numerical
calculations, and in particular whether the conductivity $\sigma$
follows a power law close to the critical energy $E_{\rm c}$. For
this purpose we will use the recursive Green's function method
\cite{CRMac80,CRMac85} to calculate the four-terminal conductance
of a disordered system for fixed disorder strength at temperature
$T=0$. Applying the finite-size scaling analysis we will compute
the critical exponent and determine the mobility edge, i.e.\ the
MIT outside the band centre.
A complementary investigation into the statistics of the energy spectrum and the
states close to the MIT can be found in Chapter \cite{CRMehS06}. An analysis of
the mathematical properties of the so-called binary-alloy or Bernoulli-Anderson
model is done in Chapter \cite{CRKarRSS06}.

\section{The Anderson Model of Localisation and its Metal-Insulator Transition}
\label{CRsec:AndersonModel}

The Anderson model \cite{CRAnd58,CRKraM93} is widely used to investigate the
phenomenon of localisation of electronic states in disordered materials. It is
based upon a tight-binding Hamiltonian in site representation
\begin{equation} \mathcal{H} = \sum_{i} \epsilon_i \ket{i}\bra{i} +
\sum_{i \neq j} t_{ij}\,\ket{i}\bra{j}\;, \label{CReq:AndersonHam}
\end{equation} where $\ket{i}$ is a localized state at site $i$ and
$t_{ij}$ are the hopping\index{hopping} parameters, which are usually restricted
to nearest neighbours. The on-site potentials $\epsilon_i$ are random numbers,
chosen according to some distribution $P(\epsilon)$ \cite{CRBulKM85,CRBulSK87}.
In what follows we take $P(\epsilon)$ to be a box distribution over the interval
$[-W/2,W/2]$, thus $W$ determines the strength of the disorder in the system.
Other distributions have also been considered \cite{CRKraM93,CRRomS03,CROhtSK99}.

For strong enough disorder, \index{disorder}$W > W_{\rm c}(E=0)$,
all states are exponentially localized and the respective wave
functions $\Psi({\bf r})$ are proportional to $\E^{(-|{\bf r}-{\bf
r}_0|/\xi)}$ for large distances $|{\bf r}-{\bf r}_0|$. Thus,
$\Psi$ is confined to a region of some finite size, which may be
described by the so-called localisation length \index{localisation
length} $\xi$. In this language extended states\index{extended
states} are characterised by $\xi\longrightarrow \infty$.
Comparing $\xi$ with the size $L$ of the system one can
distinguish between {\it strong} and {\it weak
localisation}\index{weak localisation}, for $\xi \ll L$ and
$L<\xi$, respectively\footnote{We note that the phrase {\it weak
localization} in the context of the scaling theory is often used
with a specific meaning, namely the onset of localization in large
2-dimensional samples, where the conductance decreases
logarithmically with scale \cite{CRKraM93,CRAbrALR79}.}. Here we
also assume that the phase-relaxation length $\ell_\Phi \gg L$.
Otherwise, the effective system size is determined by $\ell_\Phi$.

\begin{figure}[!htb]
\center \includegraphics[width=.7\textwidth]{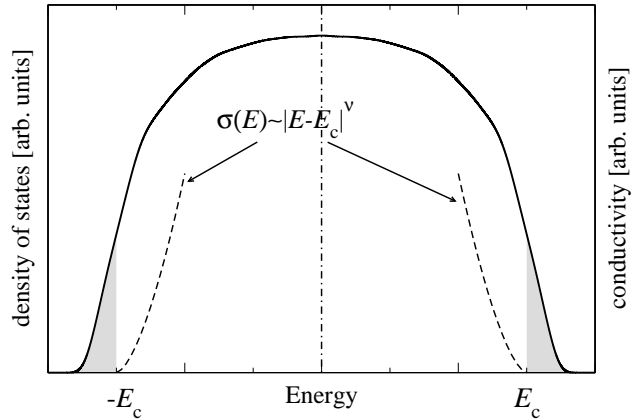}
\caption[Typical DOS of a 3D Anderson model.]{Typical DOS of a 3D
  Anderson model for fixed $W < W_{\rm c}$. The states in the grey
  regions are localized, otherwise they are extended.  The mobility
  edges are indicated at $\pm E_{\rm c}$. Also indicated is the
  power-law behaviour of $\sigma(E)$ (dashed lines) close to $\pm E_{\rm
    c}$ according to \eqref{CReq:powerlawsigma}.}
\label{CRfig:dos}
\end{figure}

It turns out that the value of the critical disorder strength \index{critical
disorder strength} $W_{\rm c}$ depends on the distribution function
$P(\varepsilon)$ and the dimension $d$ of the system. In absence of a magnetic
field and for $d{\le}2$ all states are localized\footnote{Strictly speaking, this
is only true if $\mathcal{H}$ belongs to the Gaussian orthogonal ensemble
\cite{CRAnd89}.}, i.e.\ $W_{\rm c}=0$ \cite{CRAbrALR79,CRLeeR85}. For systems
with $d=3$ the value of $W_{\rm c}$ additionally depends on the Fermi energy $E$
and the curve $W_{\rm c}(E)$ separates localized states ($W>W_{\rm c}(E)$) from
extended states ($W<W_{\rm c}(E)$) in the phase diagram
\cite{CRBulKM85,CRBulSK87,CRCaiRS99}. If instead of $E$ the disorder strength is
taken as a parameter, there will be a critical energy $E_{\rm c}(W)$
--- also called the mobility edge \index{mobility edge} --- and states with $|E|<E_{\rm c}$ are extended and those with
$|E|>E_{\rm c}$ localized yielding the same phase boundary in the ({\it E},{\it
W})-plane. At the mobility edge, states are multifractals \cite{CRMilRS97}. The
separation of localized and extended states is illustrated in Fig.\
\ref{CRfig:dos}, which shows a schematic density of states (DOS) of a
three-dimensional (3D) Anderson model.
Since for $T=0$ localized states cannot carry any electric current,
the system shows insulating behaviour, i.e. the electric conductivity
$\sigma$ vanishes for $|E| > E_{\rm c}$ or $W > W_{\rm c}$. Otherwise
the system is metallic. Therefore, the transition at the critical
point is called a {\it disorder-driven} MIT.

For the MIT in $d=3$ it was found that $\sigma$ is described by a power law at
the critical point \cite{CRKraM93},
\begin{equation} \sigma(E) = \left\{ \begin{array}{cl}
\sigma_0 \left|1-\frac{E}{E_{\rm c}} \right|^\nu, & |E| < E_{\rm c} \\
0, & |E| > E_{\rm c} \end{array}\right.\; \label{CReq:powerlawsigma}
\end{equation}
with $\nu$ being the universal critical exponent\index{universal
critical exponent} of the phase transition and $\sigma_0$ a
constant. The value of $\nu$ has been computed numerically by
various methods \cite{CRKraM93,CRSleO99a,CRMilRS00,CRMilRSU00} and
was also derived from experiments \cite{CRStuHLM93,CRWafPL99}. The
results range from $1$ to $1.6$, depending on the distribution
$P(\epsilon)$ and the computational method \cite{CRRomS03} used.

Moreover, Wegner \cite{CRWeg76} was able to show that for non-interacting
electrons the d.c.\ conductivity $\sigma$ obeys a general scaling
form\index{scaling form} close to the MIT,
\begin{equation}
        \sigma(\varepsilon,\omega) = b^{2-d}\,\sigma(b^{1/\nu} \varepsilon, b^z \omega)\;.
        \label{CReq:WegnerScaling}
\end{equation}
Here $\varepsilon$ denotes the dimensionless distance from the
critical point, $\omega$ is an external parameter such as the
frequency or the temperature, $b$ is a scaling parameter and $z$
is the dynamical exponent\index{dynamical exponent}. For
non-interacting electrons \mbox{$z=d$} \cite{CRBelK94}.
Assuming a finite conductivity for $\omega=0$, one obtains from
(\ref{CReq:WegnerScaling})
\begin{equation}
        \sigma(\varepsilon, 0) \propto \varepsilon^{\nu(d-2)},
\end{equation}
where $\varepsilon=\left|1-{E}/{E_{\rm c}} \right|$. With $d=3$
this gives (\ref{CReq:powerlawsigma}).

\section{Computational Method}
\label{CRsec:RGFM}

An approach to calculate the d.c.\ conductivity from the Anderson tight-binding
Hamiltonian (\ref{CReq:AndersonHam}) is the recursive Green's function
method\index{recursive Green's function method} \cite{CRMac80,CRMac85,CRBulSK87}.
It yields a recursion scheme for the d.c.\ conductivity tensor starting from the
Kubo-Greenwood formula\index{Kubo-Greenwood formula} \cite{CRRomMV03}. Moreover,
this method allows to compute the density of states\index{density of states} and
the localization length as well as the full set of thermoelectric kinetic
coefficients\index{thermoelectric kinetic coefficients} \cite{CRVil01}. Parallel
implementations of the method are advantageous \cite{CRCaiMRS01,CRCaiMRS02}.  The
method is therefore a companion to the more widely used transfer-matrix
method\index{transfer-matrix method} \cite{CRKraS96,CREilRS98a} or iterative
diagonalisation schemes\index{iterative diagonalisation schemes}
\cite{CRElsMMR99,CRSchMRE99}.

\subsection{Recursive Green's Function Method}

Let
$
        \mathcal{H} = \sum_{ij} H_{ij} \ket{i}\bra{j}
        \label{CReq:TBHamiltonian}
$ 
denote our hermitian tight-binding Hamiltonian. The single particle Green's
function $\mathcal{G}^\pm (z)$ is defined as \cite{CREco90}
$
        ( z^\pm - \mathcal{H} ) \mathcal{G}^\pm = \mathbbm{1}
        \label{eq:AbstractG}
$ 
where $z=E \pm \imag \gamma$ is the {\em complex energy} and the
sign of the small imaginary part $\gamma$ distinguishes between
advanced and retarded Green's functions\index{retarded Green's
functions}, $\mathcal{G}^-(E-\imag 0)$ and $\mathcal{G}^+(E+\imag
0)$, respectively \cite{CREco90}. Equivalently, $\mathcal{G}^\pm$
can be represented in the basis of the functions $\ket{i}$,
\begin{equation}
        (z^\pm \delta_{ij} - H_{ij})G^\pm_{ij} = \delta_{ij}\;,
        \label{CReq:PosBasisG}
\end{equation}
where $G^\pm_{ij}$ is the matrix element
$\bra{i}\mathcal{G}^\pm\ket{j}$. We note that for a hermitian
Hamiltonian $G^-_{ij} = (G^+_{ji})^*$.

\begin{figure}[!tbh]
        \center \includegraphics[width=.6\textwidth]{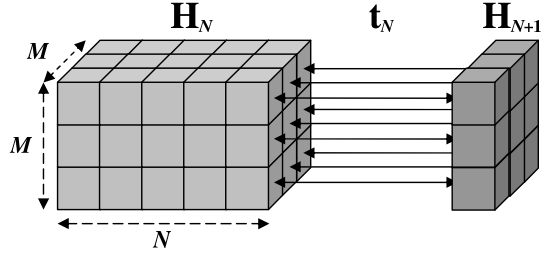}
        \caption[Scheme of the recursive Green's function method for
        a 3D system.]{Scheme of the recursive Green's function method
        for a 3D system. The new Green's function $\mathbf{G}^{(N+1)}$
        can be calculated from the old Hamiltonian $\mathbf{H}_N$ (light grey), the new slice Hamiltonian $\mathbf{H}_{N+1}$ (dark grey) and the coupling $\mathbf{t}_{N}$ (solid arrows).} \label{CRfig:RGFMScheme}
\end{figure}

If $\mathcal{H}$ contains only nearest-neighbour hopping matrix
elements, (\ref{CReq:PosBasisG}) can be simplified using a block
matrix notation. This is equivalent to considering the system as
being built up of slices or strips for 3D or 2D, respectively,
along one lattice direction. In what follows all quantities
written in bold capitals are matrices acting in the subspace of
such a slice or strip. For 2D and 3D these are matrices of size
$M\times M$ and $M^2\times M^2$, respectively, where $M$ is the
lateral extension of the system (cf.\ Fig.\
\ref{CRfig:RGFMScheme}). The left hand side of
(\ref{CReq:PosBasisG}) is then given as
\begin{gather}
    \left(  \begin{array}{cccccc}
                \ddots & \ddots & \ddots & 0 & 0 & \cdots \\
                0      & -\mathbf{H}_{i,i-1} & (z^\pm \mathbf{I} - \mathbf{H}_{i i}) & -\mathbf{H}_{i,i+1} & 0 & \cdots \\
                \cdots & 0 & -\mathbf{H}_{i+1,i} & (z^\pm \mathbf{I} - \mathbf{H}_{i+1,i+1}) & -\mathbf{H}_{i+1,i+2} & 0 \\
                \cdots & 0 & 0 & \ddots & \ddots & \ddots
            \end{array}
    \right) \times \notag \\
    \left(  \begin{array}{ccc}
                \ddots & \vdots & \ddots \\
                \cdots & \mathbf{G}^\pm_{i-1,j} & \cdots \\
                \cdots & \mathbf{G}^\pm_{i j}   & \cdots \\
                \cdots & \mathbf{G}^\pm_{i+1,j} & \cdots \\
                \ddots & \vdots & \ddots
            \end{array}
    \right)\;,
\end{gather}
where {\it i} and {\it j} now label the slices or strips. From
this expression one can easily see that (\ref{CReq:PosBasisG}) is
equivalent to
\begin{equation}
        (z^\pm \mathbf{I} - \mathbf{H}_{i i})\mathbf{G}^\pm_{i j} - \mathbf{H}_{i,i-1}\mathbf{G}^\pm_{i-1,j} - \mathbf{H}_{i,i+1}\mathbf{G}^\pm_{i+1,j} = \mathbf{I} \delta_{i j}\;.
\end{equation}
Using the hermiticity of $\mathcal{H}$ we define the hopping matrix
$\mathbf{t}_i \equiv \mathbf{H}_{i, i+1}$ (and hence
$\mathbf{t}^\dagger_i = \mathbf{H}_{i, i-1}$) connecting the $i$th and
the $(i+1)$st slice.
Now, we consider adding an additional slice to a system consisting of $N$ slices.
The Hamiltonian of this larger system can be written as \cite{CRMac85}
\begin{equation}
        \mathcal{H}^{(N+1)} \longrightarrow \mathbf{H}_{i j} + \mathbf{t}_N + \mathbf{t}^\dagger_N + \mathbf{H}_{N+1,N+1}\quad(i,j\le N).
    \end{equation}
The first and the last terms describe the uncoupled $N$-slice and the additional
$1$-slice system. Using $\mathbf{t}_N$ as an "interaction" the Green's function
$\mathbf{G}^{(N+1)}$ of the coupled system can be calculated via Dyson's equation
\cite{CRMac85,CREco90},
\begin{equation}
        \mathbf{G}^{(N+1)}_{i j} = \mathbf{G}^{(N)}_{i j} + \mathbf{G}^{(N)}_{i N} \mathbf{t}_N \mathbf{G}^{(N+1)}_{N j} \quad(i,j\le N).
\end{equation}
In particular, we have
\begin{subequations}
    \begin{align}
        \mathbf{G}^{(N+1)}_{N+1,N+1} &= \left[ z^\pm \mathbf{I} - \mathbf{H}_{N+1,N+1} - \mathbf{t}^\dagger_N \mathbf{G}^{(N)}_{N N} \mathbf{t}_N \right]^{-1} \\
        \mathbf{G}^{(N+1)}_{i j}     &= \mathbf{G}^{(N)}_{i j} + \mathbf{G}^{(N)}_{i N} \mathbf{t}_N \mathbf{G}^{(N+1)}_{N+1,N+1}  \mathbf{t}^\dagger_N \mathbf{G}^{(N)}_{N j}\quad(i,j\le N)\\
        \mathbf{G}^{(N+1)}_{i,N+1}   &= \mathbf{G}^{(N)}_{i N} \mathbf{t}_N \mathbf{G}^{(N+1)}_{N+1,N+1}\quad(i\le N) \\
        \mathbf{G}^{(N+1)}_{N+1,j}   &= \mathbf{G}^{(N+1)}_{N+1,N+1} \mathbf{t}^\dagger_N \mathbf{G}^{(N)}_{N j} \quad(j\le N)\;.
    \end{align}
    \label{CReq:RecScheme}
\end{subequations}

\noindent With (\ref{CReq:RecScheme}) the Green's function can be
obtained iteratively. Additionally, there are two kinds of
boundary conditions\index{boundary conditions} which must be
considered: across each slice and at the beginning and the end of
the stack. The first kind does not present any difficulty and
usually hard wall or periodic boundary conditions are employed.
The second kind of boundary is connected to some subtleties with
attached leads which will be addressed in Section
\ref{CRsec:LeadInfl}.

\subsection{Density of States and D.C.\ Conductivity}

The DOS \index{density of states} is given in terms of Green's function by
\cite{CREco90}
\begin{equation}
        \rho(E) = -\frac{1}{\pi \Omega} \Im \Tr \mathcal{G}^+ = -\frac{1}{\pi {\it N} {\it M}^2} \Im \sum\limits^ {\it N}_{{\it i}=1} \Tr \mathbf{G}^+_{{\it i i}}\;
\end{equation}
and the d.c.\ conductivity $\sigma$ is
\begin{equation}
        \sigma = \frac{2 e^2 \hbar}{\pi \Omega m^2} \Tr \left[ {\it p}\,\Im \mathcal{G}^+\,{\it p}\,\Im \mathcal{G}^+ \right]\;.
        \label{CReq:MomentumSigma}
\end{equation}
Here, ${\Omega}$ denotes the volume of the system and {\it m} the
electron mass. Using for the momentum the relation $p =
\frac{\imag
  m}{\hbar} [\mathcal{H},x]$ one can rewrite
(\ref{CReq:MomentumSigma}) in position representation
\begin{equation}
        \sigma = \frac{e^2 4}{h N M^2} \Tr\left\{ \gamma^2 \sum^ {\it N}_{{\it i,j}}\mathbf{G}^+_{{\it i j}}\,{\it x_j}\,\mathbf{G}^-_{{\it j i}}\,{\it
        x_i}
                                                - \imag \frac{\gamma}{2}\sum^ {\it N}_{{\it i}} (\mathbf{G}^+_{{\it i i}} - \mathbf{G}^-_{{\it i i}})\,{\it x}^2 _{\it i}\right\}\;,
\end{equation}
where $x_i$ is the position of the $i$th slice.

Starting from these relations and using the iteration scheme
(\ref{CReq:RecScheme}) one can derive recursion formul{\ae} to calculate the
properties for the $(N+1)$-slice system. The results are expressed in terms of
the following auxiliary matrices\index{auxiliary matrices}
\begin{subequations}
    \begin{align}
        \mathbf{R}_N    &= \mathbf{G}^+_{N,N}\;, \\
        \mathbf{B}_N    &= \gamma \mathbf{t}^\dagger_N \left[ \sum\limits^N_{ij} \mathbf{G}^{+(N)}_{N j} x_j (2\gamma \mathbf{G}^{- (N)}_{j i} - \imag \mathbf{I} \delta_{ij}) x_i \mathbf{G}^{+ (N)}_{i N} \right] \mathbf{t}_N\;, \\
        \mathbf{C}^+_N  &= \gamma \mathbf{t}^\dagger_N \left[ \sum\limits^N_{i=1} \mathbf{G}^{+(N)}_{N i} x_i \mathbf{G}^{- (N)}_{i N} \right] \mathbf{t}_N = (\mathbf{C}^+_N)^\dagger\;, \\
        \mathbf{C}^-_N  &= \gamma \mathbf{t}^\dagger_N \left[ \sum\limits^N_{i=1} \mathbf{G}^{-(N)}_{N i} x_i \mathbf{G}^{+ (N)}_{i N} \right] \mathbf{t}_N = (\mathbf{C}^-_N)^\dagger\;, \\
        \mathbf{F}_N    &= \mathbf{t}^\dagger_N \left[ \sum\limits^N_{i=1} \mathbf{G}^{+(N)}_{N i} \mathbf{G}^{+ (N)}_{i N} \right] \mathbf{t}_N\;.
    \end{align}
\end{subequations}
The derivation can be simplified assuming the new slice to be at
$x_{N+1}=0$. This leads, however, to corrections for the matrices
$\mathbf{B}_N$ and $\mathbf{C}^{\pm}_{N}$ because the origin of
$x_i$ has to be shifted to the position of the current slice in each
iteration step. The corrections are
\begin{subequations}
    \begin{align}
        \mathbf{B}'_N &= \mathbf{B}_N + \imag \mathbf{C}^+_N + \imag \mathbf{C}^-_N + \frac{1}{2}\mathbf{t}^\dagger_{N} (\mathbf{R}_{N}-\mathbf{R}^\dagger_{N})\mathbf{t}_{N}\;, \\
        \mathbf{C}'^{\pm}_{N} &= \mathbf{C}^\pm_N - \imag \frac{1}{2}\mathbf{t}^\dagger_{N} (\mathbf{R}_{N}-\mathbf{R}^\dagger_{N})\mathbf{t}_{N} \;.
    \end{align}
\end{subequations}
Here we have used the identity
\begin{equation}
        \gamma \sum\limits^{N}_{i=1} \mathbf{G}^+_{N i} \mathbf{G}^-_{i N} = \imag \frac{1}{2} (\mathbf{G}^+_{N N}-\mathbf{G}^-_{N N}) = \imag \frac{1}{2} (\mathbf{R}_{N}-\mathbf{R}^\dagger_{N}) = - \Im \mathbf{R}_{{\it N}}\;.
\end{equation}
The derivation of the recursion relations\index{recursion relations} is given in
Refs.\ \cite{CRMac80,CRMac85,CRCzyKM81,CRBulSK87}, it yields the following
expressions
\begin{subequations}
    \begin{align}
        s^{(N+1)}_\rho  &= s^{(N)}_\rho + \Tr\{ \mathbf{R}_{{\it N}+1} (\mathbf{F}_{\it N} + \mathbf{I}) \}\;, \\
        s^{(N+1)}_\sigma &= s^{(N)}_\sigma + \Tr\{ \Re(\mathbf{B}_{\it N} \mathbf{R}_{{\it N}+1}) + \mathbf{C}^+_{\it N} \mathbf{R}^\dagger_{{\it N}+1} \mathbf{C}^-_{\it N} \mathbf{R}_{{\it N}+1} \}\;,\\
        \mathbf{R}_{N+1}    &= \left[ z^\pm \mathbf{I} - \mathbf{H}_{N+1,N+1} - \mathbf{t}^\dagger_N \mathbf{R}_{N} \mathbf{t}_N \right]^{-1}\;, \\
        \mathbf{B}_{N+1}    &= \mathbf{t}^\dagger_{N+1} \mathbf{R}_{N+1} \left[ \mathbf{B}_{N} + 2\mathbf{C}^+_{N} \mathbf{R}^\dagger_{N+1} \mathbf{C}^-_{N} \right] \mathbf{R}_{N+1} \mathbf{t}_{N+1}\;,\\
        \mathbf{C}^+_{N+1}  &= \mathbf{t}^\dagger_{N+1} \mathbf{R}_{N+1} \mathbf{C}^+_{N} \mathbf{R}^\dagger_{N+1} \mathbf{t}_{N+1}\;,\\
        \mathbf{C}^-_{N+1}  &= \mathbf{t}^\dagger_{N+1} \mathbf{R}^\dagger_{N+1} \mathbf{C}^-_{N} \mathbf{R}_{N+1} \mathbf{t}_{N+1}\;,\\
        \mathbf{F}_{N+1}    &= \mathbf{t}^\dagger_{N+1} \mathbf{R}_{N+1}(\mathbf{F}_N + \mathbf{I})\mathbf{R}_{N+1} \mathbf{t}_{N+1}\;.
    \end{align}
    \label{CReq:RecScheme2}
\end{subequations}
The DOS and the d.c.\ conductivity are then given as
\begin{align}
        \rho^{(N+1)}(E)     &= -\frac{1}{\pi (N+1) M^2} s^{(N+1)}_\rho\;, \\
        \sigma^{(N+1)}(E)   &= \frac{e^2}{h}\frac{4}{(N+1) M^2}\;s^{(N+1)}_\sigma\;.
        \label{CReq:recRhoSigma}
\end{align}
For a comparison with the scaling arguments, we convert the
conductivity into the {\em two-terminal} conductance as
\begin{equation}\label{CReq:conducanceivity}
    g_2 = \sigma \frac{M^{2}}{L}
\end{equation}
with $L=N+1$. In distinction to the usual use of the recursive scheme which
constructs a single sample with $L \gg M$, we shall have to use many different
{\em cubic} samples with $L=M$.

We note that it is also possible to calculate the localisation
length\index{localisation length} $\xi(E)$ by the Green's function
method. The value of $\xi(E)$ is connected to the matrix
$\mathbf{G}^+_{1 N+1}$,
\begin{equation}
        \frac{1}{\xi(E)} = -\lim\limits_{\gamma\rightarrow 0} \lim\limits_{N\rightarrow \infty} \frac{1}{2 N} \ln \left| \Tr \mathbf{G}^+_{1 {\it N}} ({\it E}) \right|^2\;.
\end{equation}
The recursion relation for $\xi^{(N+1)}(E)$ is
\begin{subequations}
    \begin{align}
        \frac{1}{\xi^{(N+1)}(E)} &= -\frac{1}{N+1} s^{(N+1)}_\xi\;, \\
        s^{(N+1)}_\xi &= s^{(N)}_\xi + \ln \left| \Tr \mathbf{G}^{+({\it N}+1)}_{{\it N}+1,{\it N}+1} \right|\;.
    \end{align}
\end{subequations}

\section{Finite-Size Scaling}
\label{CRsec:fsscaling}

For finite systems there can be no singularities induced by a phase transition
and the divergences at the MIT are always rounded off \cite{CRCar96}.
Fortunately, the MIT can still be studied using a technique known as finite-size
scaling\index{finite-size scaling} \cite{CRKraM93}. Here we briefly review the
main results taking the dimensionless four-terminal
conductance\index{dimensionless conductance} $g_4$ of a large cubic sample of
size $L\times L\times L$ as an example. We note that similar scaling ideas can
also be applied to the reduced localisation length\index{reduced localisation
length} $\xi/L$.  In order to obtain $g_4$ of the disordered region only, we have
to subtract the contact resistance due to the leads. This gives
\begin{equation}
    \frac{1}{g_4} = \frac{1}{g_2} - \frac{1}{{\cal N}}\;.
    \label{CReq:SubContactRes}
\end{equation}
Here ${\cal N}={\cal N}(E)$ is the number of propagating channels at the Fermi
energy $E$ which is determined by the quantization of wave numbers in transverse
direction in the leads \cite{CRBut88b,CRCro05}.

Near the MIT one expects a one-parameter scaling\index{one-parameter scaling} law
for the dimensionless conductance \cite{CRAbrALR79,CRWeg76,CRCar96}
\begin{equation}
  g_4(L,\varepsilon,b) = \mathcal{F}\left[\frac{L}{b}, \chi(\varepsilon) b^{1/\nu}\right]\;,
        \label{CReq:SimpleScaling}
\end{equation}
where $b$ is the scale factor in the renormalisation group, $\chi$ is a
relevant scaling variable and $\nu>0$ is the critical
exponent\index{critical exponent}. The parameter
$\varepsilon$ 
measures the distance from the the mobility edge\index{mobility
edge} $E_{\rm c}$ as in (\ref{CReq:powerlawsigma}). However,
recent advances in numerical precision have shown that in addition
{\it
  corrections to scaling} due to the finite sizes of the sample need to
be taken into account so that the general scaling form\index{general
  scaling form} is
\begin{equation}
        g_4(L,\varepsilon,b) = \mathcal{F}\left[\frac{L}{b}, \chi(\varepsilon) b^{1/\nu}, \phi(\varepsilon) b^{-y}\right]\;,
        \label{CReq:GeneralScaling}
\end{equation}
where $\phi$ is an irrelevant scaling variable and $y>0$ is the corresponding
irrelevant scaling exponent\index{irrelevant scaling exponent}. The choice $b =
L$ leads to the standard scaling form\footnote{The choice of $b$ is connected to
the iteration of the renormalisation group \cite{CRCar96}.
\label{fn:RenormIteration}}
\begin{equation}
        g_4(L,\varepsilon) =
        F\left[L^{1/\nu}\chi(\varepsilon),L^{-y}\phi(\varepsilon)\right]
        \label{CReq:SlevinScaling}
\end{equation}
with $F$ being related to $\mathcal{F}$. For $E$ close to $E_{\rm
c}$ we may expand $F$ up to order $n_{\rm R}$ in its first and up to
order $n_{\rm I}$ in its second argument such that
\begin{subequations}
      \begin{align}
        g_4(L,\varepsilon)    &= \sum\limits^{n_{\rm I}}_{n'=0} \phi^{n'} L^{-n' y} F_{n'}\left(\chi L^{1/\nu}\right)\; \mbox{with}      \label{eq:ScalingExpansionA}\\
        F_{n'}(\chi L^{1/\nu}) &= \sum\limits^{n_{\rm R}}_{n=0} a_{n' n}
        \chi^n L^{n/\nu}\;.  \label{CReq:ScalingExpansionB}
      \end{align}
      \label{CReq:ScalingExpansion}
\end{subequations}
Additionally $\chi$ and $\phi$ may be expanded in terms of the small
parameter $\varepsilon$ up to orders $m_{\rm R}$ and $m_{\rm I}$,
respectively. This procedure gives
\begin{equation}
        \chi(\varepsilon) = \sum\limits^{m_{\rm R}}_{m=1} b_m \varepsilon^m, \quad \phi(\varepsilon) = \sum\limits^{m_{\rm I}}_{m'=0} c_{m'} \varepsilon^{m'}\;.
        \label{CReq:VariableExpansion}
\end{equation}
From (\ref{CReq:SlevinScaling}) and (\ref{CReq:ScalingExpansion})
one can see that a finite system size results in a systematic
shift of $g_4(L,\varepsilon=0)$ with $L$, where the direction of
the shift depends on the boundary conditions \cite{CRCar96}.
Consequently, the curves $g_4(L, \varepsilon)$ do not necessarily
intersect at the critical point\index{critical point} $\varepsilon
= 0$ as one would expect from the scaling law
(\ref{CReq:SimpleScaling}). Neglecting this effect in high
precision data will give rise to wrong values for the exponents.

Using a least-squares fit\index{least square fit} of the numerical
data to (\ref{CReq:ScalingExpansion}) and
(\ref{CReq:VariableExpansion}) allows us to extract the critical
parameters $\nu$ and $E_{\rm c}$ with high accuracy. One also
obtains the finite-size corrections\index{finite-size corrections}
and can subtract these to show the anticipated scaling behaviour.
This finite-size scaling\index{finite-size scaling} analysis has
been successfully applied to numerical calculations of the
localisation length and the conductance within the Anderson model
\cite{CRRomS03,CRSleMO01}.

\section{MIT at $E=0.5\,t$ for varying disorder}
\label{CRsec:disorder}
\subsection{Scaling of the Conductance}

We first investigate the standard case of varying disorder at a fixed energy
\cite{CRSleMO01}. We choose $t_{ij} \not = 0$ for nearest neigbours {\it i}, {\it
j} only, set $t_{ij}=t$ and $E=0.5\,t$ which is close to the band centre. We
impose hard wall boundary conditions\index{boundary conditions} in the transverse
direction. For each combination of disorder strength $W$ and system size $L$ we
generate an ensemble of $10000$ samples. The systems under investigation are
cubes of size $L\times L\times L$ for $L=4,6,8,10,12$ and $14$. For each sample
we calculate the DOS $\rho(E,L)$ and the dimensionless two-terminal conductance
$g_2$ using the recursive Green's function method explained in Section
\ref{CRsec:RGFM}. Finally we compute the average DOS $\langle \rho(E, L)
\rangle$, the {\em average} conductance\index{average conductance} $\langle
g_4(E, L) \rangle$ and the {\em typical} conductance \index{typical conductance}
$\exp \langle \ln g_4(E, L) \rangle$.

\begin{figure}[!htb]
        \center
        \includegraphics[width=.9\textwidth]{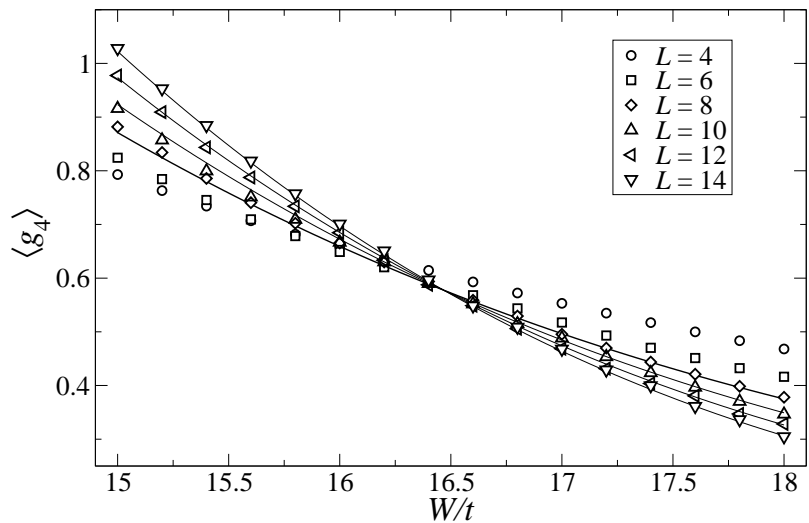}
        \caption[Average dimensionless conductance vs disorder strength
        for different system sizes at $E=0.5\,t$.]{Average dimensionless
          conductance vs disorder strength for $E=0.5\,t$. System sizes
          are given in the legend. Errors of one standard deviation are
          obtained from the ensemble average and are smaller than the
          symbol sizes. Also shown (solid lines) are fits to
          \eqref{CReq:SlevinScaling} for $L=8,10,12$ and $14$.}
        \label{CRfig:CondVsW}
\end{figure}

\begin{figure}[!htb]
        \center
        \includegraphics[width=.9\textwidth]{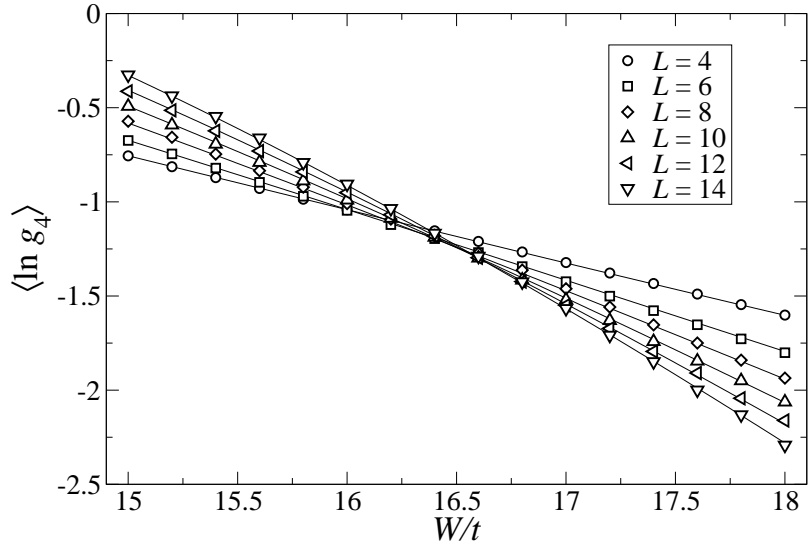}
        \caption[Logarithm of the typical dimensionless conductance vs
        disorder strength for different system sizes at
        $E=0.5\,t$.]{Logarithm of the typical dimensionless conductance
          vs disorder strength for $E=0.5\,t$. System sizes are given in
          the legend.  Errors of one standard deviation are obtained
          from the ensemble average and are smaller than the symbol
          sizes. The solid lines are fits to
          \eqref{CReq:SlevinScaling}.}  \label{CRfig:TypCondVsW}
\end{figure}

\begin{figure}[!htb]
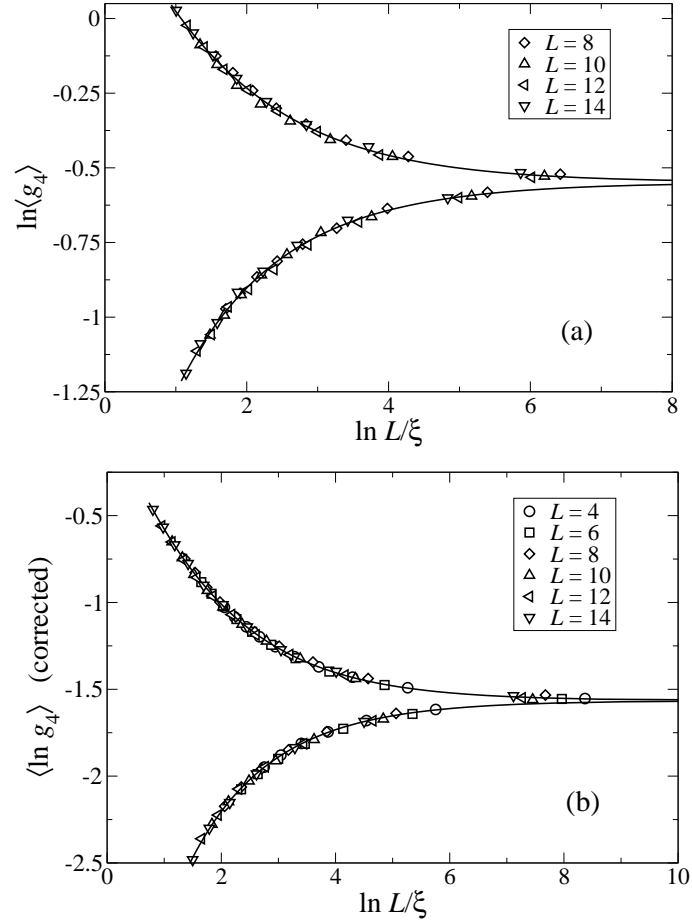

  \center
{\includegraphics[width=.75\textwidth]{Croy/fig-ScalingFunc-E05.eps}}\\[2ex]
\quad
{\includegraphics[width=.75\textwidth]{Croy/fig-ScalingFuncG-E05.eps}}
  \caption[Average and typical conductance after corrections to scaling
  are subtracted, plotted vs $L/\xi$ to show single-parameter scaling at
  $E=0.5\,t$.]{Same data as in Figs.\ \ref{CRfig:CondVsW} and
    \ref{CRfig:TypCondVsW} after corrections to scaling are subtracted,
    plotted vs $L/\xi$ to show single-parameter scaling. Different
    symbols indicate the system sizes given in the legend. The lines
    show the scaling function (\ref{CReq:SlevinScaling}).}
        \label{CRfig:ScalingFunction}
\end{figure}

The results for the different conductance averages are shown in Figs.\
\ref{CRfig:CondVsW} and \ref{CRfig:TypCondVsW} together with respective fits to
the standard scaling form (\ref{CReq:SlevinScaling}). Shown are the best fits
that we obtained for various choices of the orders of the expansions
(\ref{CReq:ScalingExpansion}, \ref{CReq:VariableExpansion}). The expansion orders
and the results for the critical exponent\index{critical exponent} and the
critical disorder\index{critical disorder} are given in Table
\ref{CRtab:FSSResults}.
\begin{table}[!tbh]
      \center
      \begin{tabular}{lcc|cccc|ccc}
        average                & $W_{\rm min}/t$ &  $W_{\rm max}/t$ & $n_{\rm R}$ & $n_{\rm I}$ &$m_{\rm R}$ &$m_{\rm I}$ & $\nu$ & $W_{\rm c}/t$ & $y$ \\
        \hline
        \hline
        $\langle g_4\rangle$     & $15.0$          &  $18.0$          & $2$         & $0$ & $2$ & $0$         & $ 1.55\pm 0.11 $  & $ 16.47\pm 0.06$  & -- \\
        $\langle \ln g_4\rangle$ & $15.0$          &  $18.0$          & $3$         & $1$ & $1$ & $0$         & $ 1.55\pm 0.18 $  & $ 16.8\pm 0.3 $ & $0.8\pm 1.0$
      \end{tabular}
      \caption{
        Best-fit estimates of the critical exponent and the critical disorder for both averages of $g_4$ using (\ref{CReq:SlevinScaling}).  The system sizes used were $L=8,10,12,14$ and $L=4,6,8,10,12,14$ for $\langle g\rangle$ and $\langle \ln g\rangle$, respectively.        For each combination of disorder strength $W$ and system size $L$ we generate an ensemble of $10000$ samples.
      }
      \label{CRtab:FSSResults}
\end{table}
In Fig.\ \ref{CRfig:ScalingFunction} we show the same data as in Figs.\
\ref{CRfig:CondVsW} and \ref{CRfig:TypCondVsW} after the corrections to scaling
have been subtracted indicating that the data points for different system sizes
fall onto a common curve with two branches as it is expected from the
one-parameter scaling theory.
The results for the conductance averages and also the critical values are in good
agreement with transfer-matrix calculations \cite{CRSleO99a,CRSleMO01}.

\subsection{Disorder Dependence of the Density of States}

\begin{figure}[!ht]
        \center \includegraphics[width=.9\textwidth]{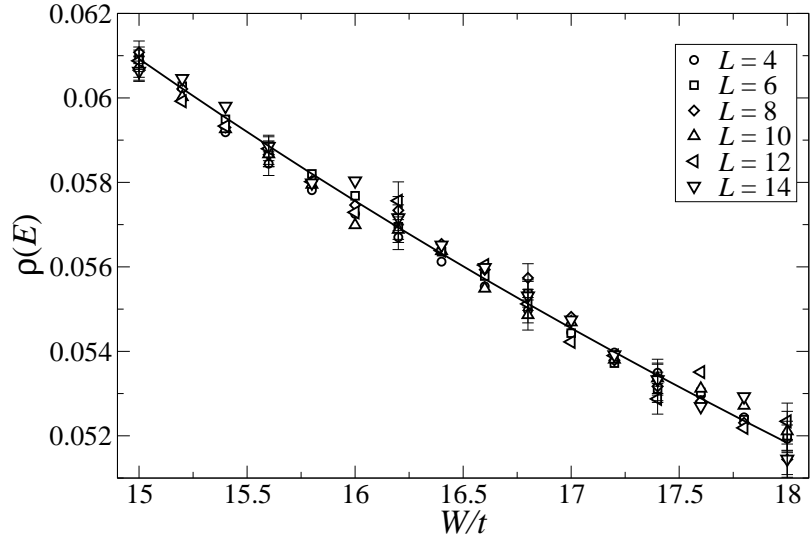}
        \caption[Density of states vs disorder strength for different
        system sizes at $E=0.5\,t$.]{Density of states vs disorder
          strength for $E=0.5\,t$ and $L=4,6,8,10,12,14$. Errors of one
          standard deviation are obtained from the ensemble average and
          shown for every 4th data point only. The solid line shows a
          fit to \eqref{CReq:DOSvsW} for $L=6$ to illustrate the
          reduction of $\rho$ with increasing disorder strength.}
        \label{CRfig:DOSVsW}
\end{figure}

The Green's function method enables us to compute the DOS of the
disordered system. It should be independent of $L$. Figure
\ref{CRfig:DOSVsW} shows the average DOS at $E=0.5\,t$ for
different system sizes. There are still some fluctuations present.
These can in principle be reduced by using larger system sizes and
increasing the number of samples. The fluctuations will be
particularly inconvenient when trying to compute $\sigma(E)$.

The reduction of the DOS with increasing disorder strength can be
understood from a simple argument. If the DOS were constant for
all energies its value would be given by the inverse of the band
width. In the Anderson model with box distribution for the on-site
energies the band width increases linearly with the disorder
strength $W$. The DOS in the Anderson model is not constant as a
function of energy, nevertheless let us assume that for energies
in the vicinity of the band centre the exact shape of the tails is
not important. Therefore,
\begin{equation}
    \rho(W) \propto \frac{1}{B + \alpha W} \;,
    \label{CReq:DOSvsW}
\end{equation}
shows a decrease of the DOS with $W$. Here {\it B} is an effective
band width taking into account that the DOS is not a constant even
for {\it W}=0. The parameter $\alpha$ allows for deviations due to
the shape of the tails. In Fig.\ \ref{CRfig:DOSVsW}, we show that
the data are indeed well described by \eqref{CReq:DOSvsW}.

\section{Influence of the Metallic Leads}
\label{CRsec:LeadInfl}

\begin{figure}[!htb]
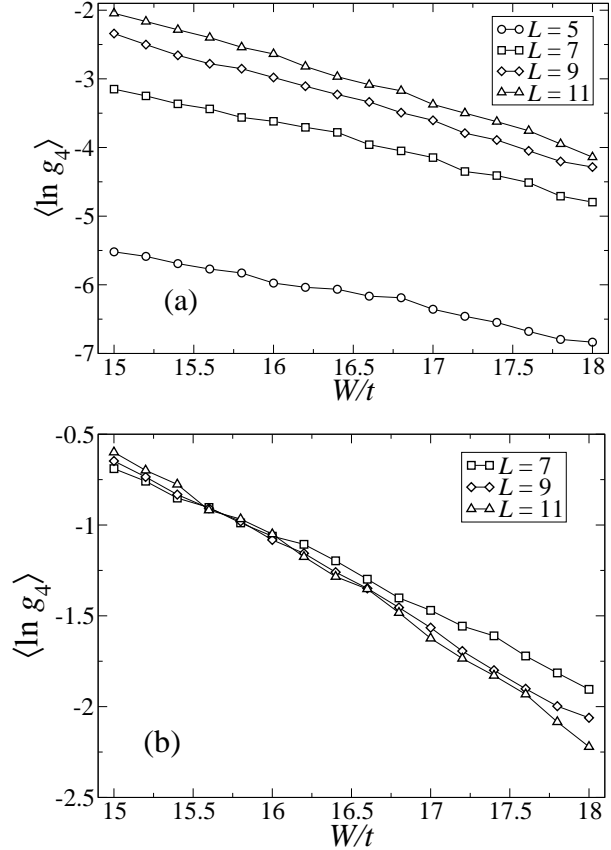

  \center 
  \quad {\includegraphics[width=.65\textwidth]{Croy/fig-DisCondG-E5-NoLS.eps}}\\[2ex]
  {\includegraphics[width=.67\textwidth]{Croy/fig-DisCondG-E5-LS.eps}}
  \caption[System size dependence of the typical conductance for fixed
  energy $E=-5\,t$.]{System size dependence of the logarithm of the
    typical conductance for fixed energy $E=-5\,t$.  Errors of one
    standard deviation are obtained from the ensemble average and are
    smaller than the symbol sizes. The lines are guides to the eye only.
    The upper plot was calculated using the metallic leads "as they are",
    i.e.\ the band centre of the leads coincides with the band centre in
    the disordered region.  In the lower plot the band centre of the
    leads was "shifted" to the respective Fermi energy.}
        \label{CRfig:E5Scaling}
\end{figure}

As mentioned in the introduction most numerical studies of the
conductance\index{conductance} have been focused on the disorder transition at or
in the vicinity of the band centre. Let us now set $E = -5\,t$ and calculate the
conductance averages as before. The results for the typical conductance are shown
in Fig.\ \ref{CRfig:E5Scaling}a. Earlier studies of the localization length
provided evidence of a phase transition around $W=16.3\, t$ although the accuracy
of the data was relatively poor \cite{CRBulSK87}. Surprisingly, in Fig.\
\ref{CRfig:E5Scaling}a there seems to be no evidence of any transition nor of any
systematic size dependence. The order of magnitude is also much smaller than in
the case of $E=0.5\,t$, although one expects the conductance at the MIT to be
roughly similar.

\begin{figure}[!htb]
        \center \includegraphics[width=.9\textwidth]{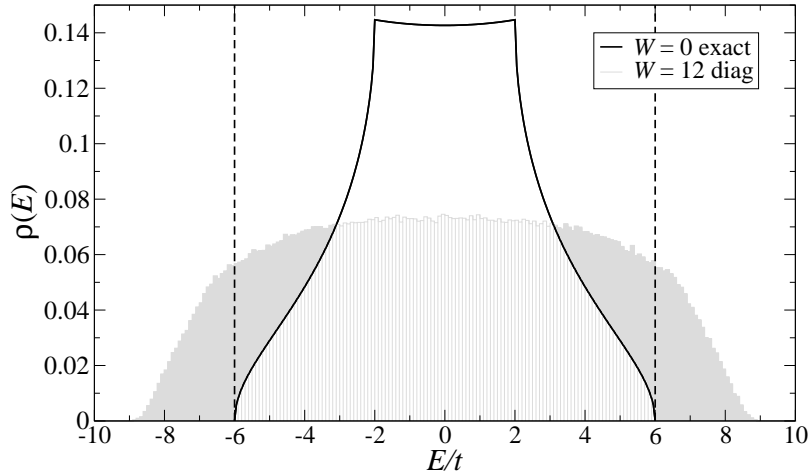}
        \caption[DODS of a clean system and a disordered system with
        $W=12 t$ and $L=21$.]{DOS of a clean system (full black line) and
          a disordered system (grey) with $W=12\,t$ and $L=21$, obtained
          from diagonalising the Hamiltonian (\ref{CReq:AndersonHam}). The
          dashed lines indicate the band edges of the ordered system.}
        \label{CRfig:LeadDOSVsE}
\end{figure}

The origin of this reduction can be understood from Fig.\ \ref{CRfig:LeadDOSVsE},
which shows the DOS of a disordered sample and a clean system (i.e.\ without
impurities and therefore without disorder) such as in the metallic
leads\index{metallic leads}. As already pointed out in Ref.\ \cite{CRMac85}, the
difference between the DOS in the leads\index{leads} and in the disordered region
may lead to false results for the transport properties. Put to an extreme, if
there are no states available at a certain energy in the leads, e.g.\ for $|E|\ge
6\,t$, there will be no transport regardless of the DOS and the conductance in
the disordered system at that energy. The DOS of the latter system becomes always
broadened by the disorder.  Therefore, using the standard setup of system and
leads, it appears problematic to investigate transport properties at energies
outside the ordered band. Additionally, for energies $3\,t \lessapprox |E| <
6\,t$ the DOS of the clean system is smaller than the disorder broadened DOS.
Thus the transport properties that crucially depend on the DOS might be also
changed in that energy range.
The problems can be overcome by shifting the energy of the
disordered region while keeping the Fermi energy\index{Fermi
energy} in the leads in the lead-band centre (or vice versa). This
is somewhat equivalent in spirit to applying a gate voltage to the
disordered region and sweeping it --- a technique similar to
MOSFET experiments. The results for the typical conductance using
this method are shown in Fig.\ \ref{CRfig:E5Scaling}. One can see
some indication of scaling behaviour and also the order of
magnitude is found to be comparable to the case of $E=0.5\,t$.
Another possibility of avoiding the DOS mismatch is choosing a larger hopping
parameter in the leads \cite{CRNik01b}, which results in a larger bandwidth, but
also a lower DOS.

\section{The MIT outside the band centre}
\label{CRsec:energy}

Knowing the difficulties involving the metallic leads and using
the "shifting technique"\index{shifting technique} explained in
the last section, we now turn our attention to the less-studied
problem of the MIT at fixed disorder. We set the disorder strength
to $W = 12\,t$ and again impose hard wall boundary conditions in
the transverse direction \cite{CRSleMO01}. We expect $E_{\rm
c}\approx 8\, t$ from the earlier studies of the localization
length \cite{CRBulSK87}. Analogous to the transition for varying
$W$ we generate for each combination of Fermi energy and system
size an ensemble of $10000$ samples (except for $L=19$ and $L=21$,
where $4000$ and $2000$ samples, respectively, were generated) and
examine the energy and size dependence of the average and the
typical conductance, $\langle g_4 \rangle$ and $\exp \langle \ln
g_4 \rangle$, respectively.

\subsection{Energy Dependence of the DOS}
\label{CRsec:EngDOS}

\begin{figure}[!htb]
            \center
            \includegraphics[width=.9\textwidth]{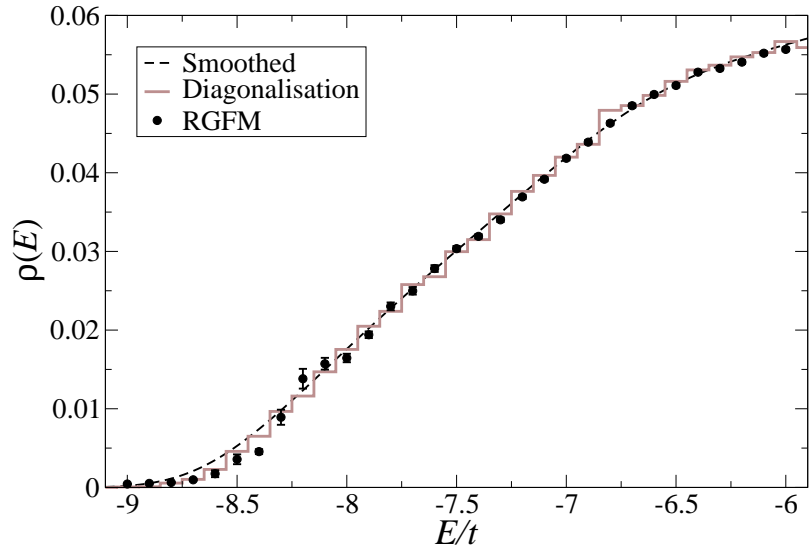}
            \caption{DOS vs energy for $W=12\,t$ obtained from the
              recursive Green's function method (circles with error bars obtained from the sample average) and from
              diagonalising the Anderson Hamiltonian (histogram), for
              $L^3=21^3=9261$. Also shown is a smoothed DOS (dashed
              line) obtained from the diagonalisation data using a
              Bezier spline\index{Bezier spline}.}
            \label{CRfig:DOSVsEDiagRGFM}
\end{figure}

\begin{figure}[!htb]
            \center
            \includegraphics[width=.9\textwidth]{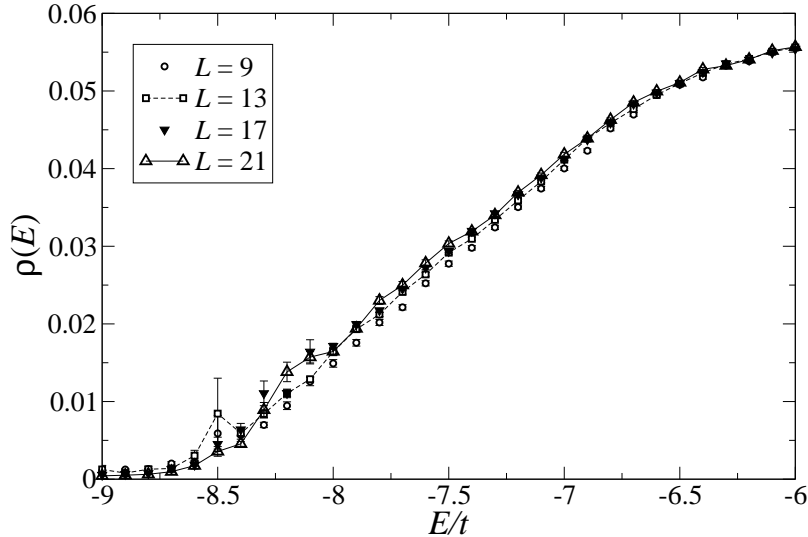}
            \caption{DOS vs energy for different system sizes and
              $W=12\,t$ calculated with the recursive Green's function
              method. The data are averaged over $10000$ disorder
              configurations except for $L=21$ when $2000$ samples have
              been used. The lines are guides to the eye only. Error bars are obtained from the sample average.}
            \label{CRfig:DOSVsE}
\end{figure}

Before looking at the scaling behaviour of the conductance we have
to make sure that the "shifting technique"\index{shifting technique}
indeed gives the right DOS outside the ordered band. Additionally,
we have to check the average DOS for being independent of the
system size.
In Fig.\ \ref{CRfig:DOSVsEDiagRGFM} we show the DOS obtained from diagonalisation
of the Anderson Hamiltonian with $30$ configurations (using standard
LAPACK\index{LAPACK} subroutines) and the Green's function calculations. The
Green's function data agree very well with the diagonalisation results, although
there are still bumps around $E=-8.2\, t$ for small DOS values.
The average DOS for different system sizes is shown in Fig.\ \ref{CRfig:DOSVsE}.
For large energies the DOS is nearly independent of the system size.
However, close to the band edge one can see fluctuations because in the
tails there are only few states and thus many more samples are necessary
to obtain a smooth DOS.

\subsection{Scaling Behaviour of the Conductance}

\begin{figure}[!htb]
            \center \includegraphics[width=.7\textwidth]{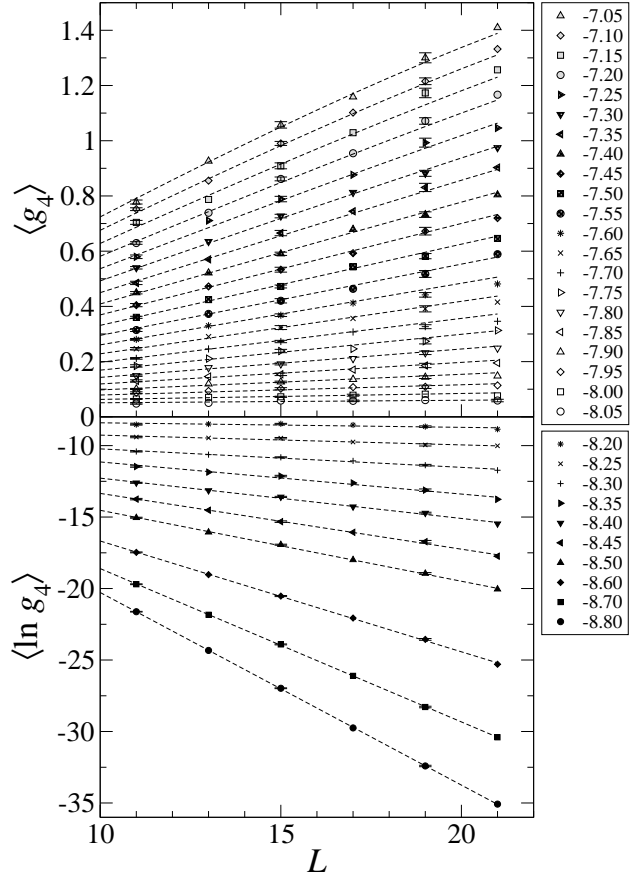}
            \caption[System size dependence of the 4-point conductance
            averages $\langle g_4 \rangle$ and $\langle \ln g_4 \rangle$
            for $W=12\,t$ and Fermi energies as given in the
            legend.]{System size dependence of the 4-point conductance
              averages $\langle g_4 \rangle$ and $\langle \ln g_4
              \rangle$ for $W=12\,t$ and Fermi energies as given in the
              legend. Error bars are obtained from the ensemble average and shown for every second
              $L$. The dashed lines in the metallic regime indicate the fit
              result to (\ref{CReq:ScalingAC}) using the parameters
              of Table \ref{CRtab:FSSEnergyResults}. In the insulating
              regime, linear functions for $\langle \ln g_4 \rangle =
              -L/\xi +c$ have been used for fitting.}
            \label{CRfig:ConductanceVsL}
\end{figure}

\begin{figure}[!htb]
            \center
            \includegraphics[width=.8\textwidth]{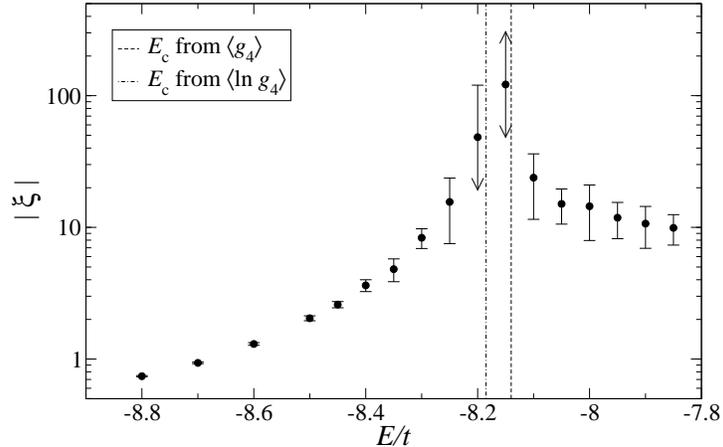}
            \caption{Localisation length (for $E < E_{\rm c}$) and
              correlation length (for $E>E_{\rm c}$) vs energy obtained
              from a linear fit to $\langle \ln g_4 \rangle = \mp L/\xi
              + {\rm const.}$, respectively. The error bars close to the
              transition have been truncated (arrows), because they
              extend beyond the plot boundaries.}
            \label{CRfig:LocLength}
\end{figure}

The size dependence of the average and the typical conductance is shown in Fig.\
\ref{CRfig:ConductanceVsL}. We find that for $E/t \le -8.2$ the typical
conductance is proportional to the system size $L$ and the constant of
proportionality is negative. This corresponds to an exponential decay of the
conductance with increasing $L$ and is characteristic for {\it insulating}
behaviour. Moreover, the constant of proportionality is the localisation length
\index{localisation length} $\xi$. We find that $\xi(E)$ diverges at some energy,
which indicates a phase transition. This energy dependence of $\xi$ is shown in
Fig.\ \ref{CRfig:LocLength}.

For $E/t\geq -8.05$, $\langle g_4 \rangle$ is
proportional to $L$. This indicates the {\it metallic} regime and the
slope of $\langle g_4 \rangle$ vs $L$ is related to the d.c.\
conductivity.
We fit the data in the respective regimes to the standard scaling form
(\ref{CReq:SlevinScaling}). The results for the critical exponent\index{critical
exponent} and the mobility edge\index{mobility edge} are given in Table\
\ref{CRtab:FSSEnergyResults}. The obtained values from both averages, $\langle
g_4\rangle$ and $\langle \ln g_4\rangle$, are consistent. The average value of
$\nu= 1.59 \pm 0.18$ is in agreement with results for conductance scaling at
$E/t=0.5$ and transfer-matrix calculations \cite{CRSleO99a,CRSleMO01}.
\begin{table}[!tbh]
      \center
      \begin{tabular}{lcccccc}
        average                & $E_{\rm min}/t$ &  $E_{\rm max}/t$ & $n_{\rm R}$ & $m_{\rm R}$ & $\nu$            & $E_{\rm c}/t$ \\
        \hline
        \hline
        $\langle g_4\rangle$     & $-8.2$          &  $-7.4$          & $3$         & $2$         & $1.60 \pm 0.18$  & $-8.14  \pm 0.02$ \\
        $\langle \ln g_4\rangle$ & $-8.8$          &  $-7.85$         & $3$         & $2$         & $1.58 \pm 0.06$  & $-8.185 \pm 0.012$
      \end{tabular}
      \caption[Best-fit estimates of the critical exponent and the critical disorder for both averages of $g_4$ at $W=12\,t$.]{Best-fit estimates of the critical exponent and the mobility edge for both averages of $g_4$ using (\ref{CReq:SlevinScaling}) with $n_{\rm R}=m_{\rm I}=0$. The system sizes used are  $L=11,13,15,17,19,21$.}
      \label{CRtab:FSSEnergyResults}
\end{table}

\subsection{Calculation of the D.C.\ Conductivity}
\label{CRsec:NumConductivity}

Let us now compute the d.c.\ conductivity from the conductance $\langle g_4(E, L)
\rangle$. From Ohm's law, one naively expects the macroscopic conductivity to be
the ratio of $\langle g_4(E, L) \rangle$ and $L$. There are, however, several
complications. First, the mechanism of weak localisation gives rise to
corrections to the classical behaviour for $g_4\gg 1$. Second, it is a priori not
known if this relation still holds in the critical regime. And third, the
expansion (\ref{CReq:ScalingExpansion}) does not yield a behaviour of the form
$g_4\propto \varepsilon^\nu$.

In order to check our data for consistence with the anticipated power law
(\ref{CReq:powerlawsigma}) for the conductivity $\sigma(E)$ in the critical
regime, we assume the following scaling law for the conductance,
\begin{equation}
            \langle g_4\rangle = f( \chi^\nu L )\;,
    \label{CReq:ScalingAC}
\end{equation}
which results from setting $b=\chi^{-\nu}$ in
(\ref{CReq:SimpleScaling}).  Due to the relatively large error
bars of $\langle g_4\rangle$ at the MIT as shown in Fig.\
\ref{CRfig:ConductanceVsL}, we might as well neglect the
irrelevant scaling variable.
Then we expand $f$ as a Taylor series up to order $n_{\rm R}$ and
$\chi$ in terms of $\varepsilon$ up to order $m_{\rm R}$ in
analogy to (\ref{CReq:ScalingExpansionB}) and
(\ref{CReq:VariableExpansion}).
The best fit to our data is determined by minimising the $\chi^2$
statistic. Using $n_{\rm R}=3$ and $m_{\rm R}=2$ we obtain for the
critical values, $\nu = 1.58 \pm 0.18$ and $E_{\rm c}/t = -8.12 \pm
0.03$. These values are consistent with our previous fits. The linear
term of the expansion of $f$ corresponds to the conductivity close to
the MIT. To estimate the quality of this procedure we also calculate the
conductivity from the slope of a linear fit to $\langle g_4\rangle$
throughout the metallic regime, and from the ratio $\langle
g_4\rangle/L$ as well.

\begin{figure}[!htb]
            \center
            \includegraphics[width=.9\textwidth]{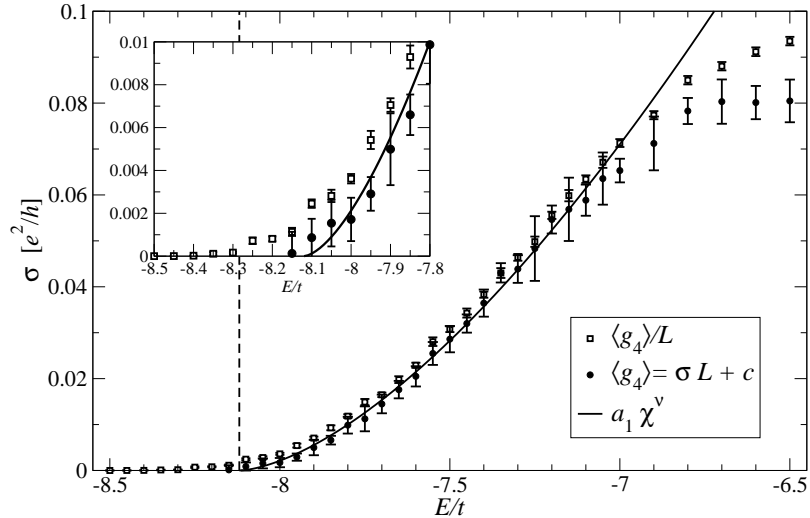}
            \caption[Conductivity $\sigma$ vs energy computed from
            $\langle g_4\rangle/L$ , a linear fit with $\langle
            g_4\rangle = \sigma L +{\rm const.}$ and a scaling form.
            ]{Conductivity $\sigma$ vs energy computed from $\langle
              g_4\rangle/L$ for $L=21$ ($\square$), a linear fit with
              $\langle g_4\rangle = \sigma L +{\rm const.}$ ($\bullet$)
              and a fit to the scaling law (\ref{CReq:ScalingAC}) (solid
              line). The dashed line indicates $E_{\rm c}/t= -8.12$.
              Error bars of $\langle g_4\rangle/L$ represent the
              error-of-mean obtained from an ensemble average and are shown for every third $E$ value. The
              dashed line indicates the position of $E_{\rm c}$ and the
              inset shows the region close to $E_{\rm c}$ in more
              detail.}
            \label{CRfig:ConductivityVsE}
\end{figure}

The resulting estimates of the conductivity are shown in Fig.\
\ref{CRfig:ConductivityVsE}.
We find that the power law is in good agreement with the
conductivity obtained from the linear fit to $\langle g_4\rangle =
\sigma L +{\rm const.}$ for $E\le -7\,t$. In this range it is also
consistent with the ratio of $\langle g_4\rangle$ and $L$ for the
largest system computed \mbox{($L=21$)}. Deviations occur for
energies close to the MIT and for $E>-7\,t$. In the critical
regime one can argue that for finite systems the conductance will
always be larger than zero in the insulating regime because the
localisation length becomes eventually larger than the system
size.
%

\section{Conclusions}
\label{CRsec:conclusions}

We computed the conductance at $T=0$ and the DOS of the 3D Anderson
model of localisation. These properties were obtained from the recursive
Green's function method in which semi-infinite metallic leads at both
ends of the system were taken into account.

We demonstrated how the difference in the DOS between the disordered region and
the metallic leads has a significant influence on the results for the electronic
properties at energies outside the band centre.  This poses a big problem for the
investigation of the MIT outside the band centre. We showed that by shifting the
energy levels in the disordered region the mismatch can be reduced. In this case
the average conductance and the typical conductance were found to be consistent
with the one-parameter scaling theory at the transition at $E_{\rm c}\neq 0$.
Using a finite-size-scaling analysis of the energy dependence of both conductance
averages we obtained an average critical exponent $\nu= 1.59 \pm 0.18$, which is
in accordance with results for conductance scaling at $E/t=0.5$, transfer-matrix
calculations \cite{CRSleO99a,CRSleMO01,CRCaiRS99,CRMilRSU00} and diagonalisation
studies \cite{CRNdaRS02a,CRMilRS00}.  However, a thorough investigation of the
influence of the leads is still lacking. It would also be interesting to see if
these effects can be related to studies of 1D multichannel systems with
impurities \cite{CRBoeLR00}.

We calculated the d.c.\ conductivity from the system-size dependence of the
average conductance and found it consistent with a power-law form at the MIT
\cite{CRCaiNRS01}. This strongly supports previous analytical and numerical
calculations of thermoelectric properties reviewed in Ref.\ \cite{CRRomS03}.
Similar results for topologically disordered Anderson models
\cite{CRZhoGRS98,CRGriRS98,CRGriRSZ00}, random-hopping models
\cite{CREilRS98a,CREilRS98b,CRBisCRS00,CREilRS01} and the interplay of disorder
and many-body interaction \cite{CRRomP95,CRLeaRS99,CRRomSV01,CRSchRS02a} have
been reported elsewhere.

\section{Acknowledgments}
\label{sec:acknow}
We thankfully acknowledge fruitful and localized discussions with P.\
Cain, F.\ Milde, M.L.\ Ndawana and C.\ de los Reyes Villagonzalo.


\begin{thebibliography}{10}

\bibitem{CRAnd58}
P.~W. Anderson.
\newblock Absence of diffusion in certain random lattices.
\newblock {\em Phys. Rev.}, 109:1492--1505, 1958.

\bibitem{CRKraM93}
B.~Kramer and A.~MacKinnon.
\newblock Localization: theory and experiment.
\newblock {\em Rep. Prog. Phys.}, 56:1469--1564, 1993.

\bibitem{CRRomS03}
R.~A. {R\"{o}mer} and M.~Schreiber. Numerical investigations of
scaling at the Anderson transition.
\newblock {In T. Brandes and S. Kettemann, editors, \em The Anderson Transition and its Ramifications --- Localisation,
Quantum Interference, and Interactions}, volume 630 of {\em
Lecture Notes in Physics}, pages 3--19.
\newblock Springer, Berlin, 2003.

\bibitem{CRPlyRS03}
I.~Plyushchay, R.~A. {R\"{o}mer}, and M.~Schreiber.
\newblock The three-dimensional Anderson model of localization with binary
  random potential.
\newblock {\em Phys. Rev. B}, 68:064201, 2003.

\bibitem{CREndB94}
J.~E. Enderby and A.~C. Barnes.
\newblock Electron transport at the {Anderson} transition.
\newblock {\em Phys. Rev. B}, 49:5062, 1994.

\bibitem{CRVilRS99a}
C.~Villagonzalo, R.~A. {R\"{o}mer}, and M.~Schreiber.
\newblock Thermoelectric transport properties in disordered systems near the
  {Anderson} transition.
\newblock {\em Eur. Phys. J. B}, 12:179--189, 1999.
\newblock {ArXiv}: cond-mat/9904362.

\bibitem{CRAbrALR79}
E.~Abrahams, P.~W. Anderson, D.~C. Licciardello, and T.~V. Ramakrishnan.
\newblock Scaling theory of localization: absence of quantum diffusion in two
  dimensions.
\newblock {\em Phys. Rev. Lett.}, 42:673--676, 1979.

\bibitem{CRLeeR85}
P.~A. Lee and T.~V. Ramakrishnan.
\newblock Disordered electronic systems.
\newblock {\em Rev. Mod. Phys.}, 57:287--337, 1985.

\bibitem{CRSleO99a}
K.~Slevin and T.~Ohtsuki.
\newblock Corrections to scaling at the {Anderson} transition.
\newblock {\em Phys. Rev. Lett.}, 82:382--385, 1999.
\newblock {ArXiv}: cond-mat/9812065.

\bibitem{CRMilRS00}
F.~Milde, R.~A. {R\"{o}mer}, and M.~Schreiber.
\newblock Energy-level statistics at the metal-insulator transition in
  anisotropic systems.
\newblock {\em Phys. Rev. B}, 61:6028--6035, 2000.

\bibitem{CRMilRSU00}
F.~Milde, R.~A. {R\"{o}mer}, M.~Schreiber, and V.~Uski.
\newblock Critical properties of the metal-insulator transition in anisotropic
  systems.
\newblock {\em Eur. Phys. J. B}, 15:685--690, 2000.
\newblock {ArXiv}: cond-mat/9911029.

\bibitem{CRNdaRS02a}
M.~L. Ndawana, R.~A. {R\"{o}mer}, and M.~Schreiber.
\newblock Finite-size scaling of the level compressibility at the {Anderson}
  transition.
\newblock {\em Eur. Phys. J. B}, 27:399--407, 2002.

\bibitem{CRNdaRS04}
M.~L. Ndawana, R.~A. {R\"{o}mer}, and M.~Schreiber.
\newblock Effects of scale-free disorder on the {Anderson} metal-insulator
  transition.
\newblock {\em Europhys. Lett.}, 68:678--684, 2004.

\bibitem{CRSleMO01}
K.~Slevin, P.~Marko\u{s}, and T.~Ohtsuki.
\newblock Reconciling conductance fluctuations and the scaling theory of
  localization.
\newblock {\em Phys. Rev. Lett.}, 86:3594--3597, 2001.

\bibitem{CRBraHMM01}
D.~Braun, E.~Hofstetter, G.~Montambaux, and A.~MacKinnon.
\newblock Boundary conditions, the critiical conductance distribution, and
  one-parameter scaling.
\newblock {\em Phys. Rev. B}, 64:155107, 2001.

\bibitem{CRVilRS99b}
C.~Villagonzalo, R.~A. {R\"{o}mer}, and M.~Schreiber.
\newblock Transport properties near the {Anderson} transition.
\newblock {\em {Ann. Phys. (Leipzig)}}, 8:SI-269--SI-272, 1999.
\newblock {ArXiv}: cond-mat/9908218.

\bibitem{CRVilRSM00a}
C.~Villagonzalo, R.~A. {R\"{o}mer}, M.~Schreiber, and A.~MacKinnon.
\newblock Behavior of the thermopower in amorphous materials at the
  metal-insulator transition.
\newblock {\em Phys. Rev. B}, 62:16446--16452, 2000.

\bibitem{CRMac80}
A.~MacKinnon.
\newblock The conductivity of the one-dimensional disordered Anderson model: a
  new numerical method.
\newblock {\em J. Phys.: Condens. Matter}, 13:L1031--L1034, 1980.

\bibitem{CRMac85}
A.~MacKinnon.
\newblock The calculation of transport properties and density of states of
  disordered solids.
\newblock {\em Z. Phys. B}, 59:385--390, 1985.

\bibitem{CRMehS06}
B.~Mehlig and M.~Schreiber. Energy-level and wave-function
statistics in the Anderson model of localization.
\newblock {In K.H. Hoffmann and A. Meyer, editors, \em Parallel Algorithms and Cluster Computing - Implementations,
Algorithms, and Applications},
\newblock {\em Lecture Notes in Computational Science and Engineering}. Springer,
Berlin, 2006.

\bibitem{CRKarRSS06}
P.~Karmann, R.~A. {R\"{o}mer}, M.~Schreiber, and P.~Stollmann.
Fine structure of the integrated density of states for
Bernoulli-Anderson models.
\newblock {In K.H. Hoffmann, and A. Meyer, editors, \em Parallel Algorithms and Cluster Computing - Implementations,
  Algorithms, and Applications},
\newblock {\em Lecture Notes in Computational Science and Engineering}. Springer,
  Berlin, 2006.

\bibitem{CRBulKM85}
B.~Bulka, B.~Kramer, and A.~MacKinnon.
\newblock Mobility edge in the three dimensional {Anderson} model.
\newblock {\em Z. Phys. B}, 60:13--17, 1985.

\bibitem{CRBulSK87}
B.~Bulka, M.~Schreiber, and B.~Kramer.
\newblock Localization, quantum interference, and the metal-insulator
  transition.
\newblock {\em Z. Phys. B}, 66:21, 1987.

\bibitem{CROhtSK99}
T.~Ohtsuki, K.~Slevin, and T.~Kawarabayashi.
\newblock Review on recent progress on numerical studies of the {Anderson}
  transition.
\newblock {\em {Ann. Phys. (Leipzig)}}, 8:655--664, 1999.
\newblock {ArXiv}: cond-mat/9911213.

\bibitem{CRAnd89}
T.~Ando.
\newblock {\em Phys. Rev. B}, 40:5325, 1989.

\bibitem{CRCaiRS99}
P.~Cain, R.~A. {R\"{o}mer}, and M.~Schreiber.
\newblock Phase diagram of the three-dimensional {Anderson} model of
  localization with random hopping.
\newblock {\em {Ann. Phys. (Leipzig)}}, 8:SI-33--SI-38, 1999.
\newblock {ArXiv}: cond-mat/9908255.

\bibitem{CRMilRS97}
F.~Milde, R.~A. {R\"{o}mer}, and M.~Schreiber.
\newblock Multifractal analysis of the metal-insulator transition in
  anisotropic systems.
\newblock {\em Phys. Rev. B}, 55:9463--9469, 1997.

\bibitem{CRStuHLM93}
H.~Stupp, M.~Hornung, M.~Lakner, O.~Madel, and H.~v.~{L\"{o}hneysen}.
\newblock Possible solution of the conductivity exponent puzzle for the
  metal-insulator transition in heavily doped uncompensated semiconductors.
\newblock {\em Phys. Rev. Lett.}, 71:2634--2637, 1993.

\bibitem{CRWafPL99}
S.~Waffenschmidt, C.~Pfleiderer, and H.~v.~{L\"{o}hneysen}.
\newblock Critical behavior of the conductivity of {Si:P} at the
  metal-insulator transition under uniaxial stress.
\newblock {\em Phys. Rev. Lett.}, 83:3005--3008, 1999.
\newblock {ArXiv}: cond-mat/9905297.

\bibitem{CRWeg76}
F.~Wegner.
\newblock Electrons in disordered systems. Scaling near the mobility edge.
\newblock {\em Z. Phys. B}, 25:327--337, 1976.

\bibitem{CRBelK94}
D.~Belitz and T.~R. Kirkpatrick.
\newblock The {Anderson}-{Mott} transition.
\newblock {\em Rev. Mod. Phys.}, 66:261--380, 1994.

\bibitem{CRRomMV03}
R.~A. {R\"{o}mer}, C.~Villagonzalo, and A.~MacKinnon.
\newblock Thermoelectric properties of disordered systems.
\newblock {\em J. Phys. Soc. Japan}, 72:167--168, 2002.
\newblock Suppl. A.

\bibitem{CRVil01}
C.~Villagonzalo.
\newblock {\em Thermoelectric Transport at the Metal-Insulator Transition in
  Disordered Systems}.
\newblock PhD thesis, Chemnitz University of Technology, 2001.

\bibitem{CRCaiMRS01}
P.~Cain, F.~Milde, R.A. {R\"{o}mer}, and M.~Schreiber.
Applications of cluster computing for the Anderson model of
localization.
\newblock {In S.G. Pandalai, editor, \em Recent Research Developments in
Physics}, volume~2, pages 171--184.
\newblock Transworld Research Network, Trivandrum, India, 2001.

\bibitem{CRCaiMRS02}
P.~Cain, F.~Milde, R.~A. {R\"{o}mer}, and M.~Schreiber.
\newblock Use of cluster computing for the Anderson model of localization.
\newblock {\em Comp. Phys. Comm.}, 147:246--250, 2002.

\bibitem{CRKraS96}
B.~Kramer and M.~Schreiber.
\newblock Transfer-matrix methods and finite-size scaling for disordered
  systems.
\newblock In K.~H. Hoffmann and M.~Schreiber, editors, {\em Computational
  Physics}, pages 166--188, Springer, Berlin, 1996.

\bibitem{CREilRS98a}
A.~Eilmes, R.~A. {{R\"{o}mer}}, and M.~Schreiber.
\newblock The two-dimensional {Anderson} model of localization with random
  hopping.
\newblock {\em Eur. Phys. J. B}, 1:29--38, 1998.

\bibitem{CRElsMMR99}
U.~Elsner, V.~Mehrmann, F.~Milde, R.~A. {R\"{o}mer}, and M.~Schreiber.
\newblock The {Anderson} model of localization: a challenge for modern
  eigenvalue methods.
\newblock {\em SIAM J. Sci. Comp.}, 20:2089--2102, 1999.
\newblock {ArXiv}: physics/9802009.

\bibitem{CRSchMRE99}
M.~Schreiber, F.~Milde, R.~A. {R\"{o}mer}, U.~Elsner, and V.~Mehrmann.
\newblock Electronic states in the {Anderson} model of localization:
  benchmarking eigenvalue algorithms.
\newblock {\em Comp. Phys. Comm.}, 121--122:517--523, 1999.

\bibitem{CREco90}
E.~N. Economou.
\newblock {\em Green's Functions in Quantum Physics}.
\newblock Springer-Verlag, Berlin, 1990.

\bibitem{CRCzyKM81}
G.~Czycholl, B.~Kramer, and A.~MacKinnon.
\newblock Conductivity and localization of electron states in one dimensional
  disordered systems: further numerical results.
\newblock {\em Z. Phys. B}, 43:5--11, 1981.

\bibitem{CRCar96}
J.~L. Cardy.
\newblock {\em Scaling and Renormalization in Statistical Physics}.
\newblock Cambridge University Press, Cambridge, 1996.

\bibitem{CRBut88b}
M.~B\"{u}ttiker.
\newblock Absence of backscattering in the quantum Hall effect in multiprobe
  conductors.
\newblock {\em Phys. Rev. B}, 38:9375, 1988.

\bibitem{CRCro05}
A. Croy.
\newblock Thermoelectric properties of disordered systems.
\newblock {M.Sc.} thesis, University of Warwick, Coventry, United Kindgom,
  2005.

\bibitem{CRNik01b}
B.~K. Nikoli\'{c}.
\newblock Statistical properties of eigenstates in three-dimensional mesoscopic
  systems with off-diagonal or diagonal disorder.
\newblock {\em Phys. Rev. B}, 64:14203, 2001.

\bibitem{CRBoeLR00}
D.~Boese, M.~Lischka, and L.E. Reichl.
\newblock Scaling behaviour in a quantum wire with scatterers.
\newblock {\em Phys. Rev. B}, 62:16933, 2000.

\bibitem{CRCaiNRS01}
P.~Cain, M.~L. Ndawana, R.~A. {R\"{o}mer}, and M.~Schreiber.
\newblock The critical exponent of the localization length at the Anderson
  transition in {$3D$} disordered systems is larger than $1$.
\newblock 2001.
\newblock {ArXiv}: cond-mat/0106005.

\bibitem{CRZhoGRS98}
J.~X. Zhong, U.~Grimm, R.~A. {R\"{o}mer}, and M.~Schreiber.
\newblock Level spacing distributions of planar quasiperiodic tight-binding
  models.
\newblock {\em Phys. Rev. Lett.}, 80:3996--3999, 1998.

\bibitem{CRGriRS98}
U.~Grimm, R.~A. {R\"{o}mer}, and G.~Schliecker.
\newblock Electronic states in topologically disordered systems.
\newblock {\em {Ann. Phys. (Leipzig)}}, 7:389--393, 1998.

\bibitem{CRGriRSZ00}
U.~Grimm, R.~A. {R\"{o}mer}, M.~Schreiber, and J.~X. Zhong.
\newblock Universal level-spacing statistics in quasiperiodic tight-binding
  models.
\newblock {\em Mat. Sci. Eng. A}, 294-296:564, 2000.
\newblock {ArXiv}: cond-mat/9908063.

\bibitem{CREilRS98b}
A.~Eilmes, R.~A. {R\"{o}mer}, and M.~Schreiber.
\newblock Critical behavior in the two-dimensional {Anderson} model of
  localization with random hopping.
\newblock {\em phys. stat. sol. (b)}, 205:229--232, 1998.

\bibitem{CRBisCRS00}
P.~Biswas, P.~Cain, R.~A. {R\"{o}mer}, and M.~Schreiber.
\newblock Off-diagonal disorder in the {Anderson} model of localization.
\newblock {\em phys. stat. sol. (b)}, 218:205--209, 2000.
\newblock {ArXiv}: cond-mat/0001315.

\bibitem{CREilRS01}
A.~Eilmes, R.~A. {R\"{o}mer}, and M.~Schreiber.
\newblock Localization properties of two interacting particles in a
  quasi-periodic potential with a metal-insulator transition.
\newblock {\em Eur. Phys. J. B}, 23:229--234, 2001.
\newblock {ArXiv}: cond-mat/0106603.

\bibitem{CRRomP95}
R.~A. {R\"{o}mer} and A.~Punnoose.
\newblock Enhanced charge and spin currents in the one-dimensional disordered
  mesoscopic {Hubbard} ring.
\newblock {\em Phys. Rev. B}, 52:14809--14817, 1995.

\bibitem{CRLeaRS99}
M.~Leadbeater, R.~A. {R\"{o}mer}, and M.~Schreiber.
\newblock Interaction-dependent enhancement of the localisation length for two
  interacting particles in a one-dimensional random potential.
\newblock {\em Eur. Phys. J. B}, 8:643--652, 1999.

\bibitem{CRRomSV01}
R.~A. {R\"{o}mer}, M.~Schreiber, and T.~Vojta.
\newblock Disorder and two-particle interaction in low-dimensional quantum
  systems.
\newblock {\em Physica E}, 9:397--404, 2001.

\bibitem{CRSchRS02a}
C.~Schuster, R.~A. {R\"{o}mer}, and M.~Schreiber.
\newblock Interacting particles at a metal-insulator transition.
\newblock {\em Phys. Rev. B}, 65:115114--7, 2002.

\end{thebibliography}
%


\printindex
\end{document}